\documentclass[]{spie}

\usepackage{graphicx}
\usepackage{amsmath}
\usepackage{url}
\usepackage{relsize}
\usepackage{subfigure}
\usepackage{rotate}

\newcommand\cpp{C\nolinebreak[4]\hspace{-.05em}\raisebox{.6ex}{\relsize{-3}{\textbf{++}}}}

\title{SuperSpec: design concept and circuit simulations}

\author{Attila Kov\'{a}cs\supit{a,b}, Peter S.\ Barry\supit{c}, Charles M.\ Bradford\supit{d}, Goutam Chattopadhyay\supit{d}, Peter Day,\supit{d} Simon Doyle\supit{c}, Steve Hailey-Dunsheath\supit{a}, Matthew Hollister\supit{a,d}, Christopher McKenney\supit{a}, Henry G.\ LeDuc\supit{d}, Nuria Llombart\supit{e}, Daniel P.\ Marrone\supit{f}, Philip Mauskopf\supit{c,f}, Roger O'Brient\supit{a}, Stephen Padin\supit{a}, Loren J.\ Swenson\supit{a,d}, and Jonas Zmuidzinas\supit{a,d}
{\small
\skiplinehalf
\supit{a} California Institute of Technology 301-17, 1200 E.~California Blvd, Pasadena, CA 91125, USA
\skiplinehalf
\supit{b} Institute for Astrophysics, University of Minnesota, 116 Church St SE, Minneapolis, MN 55455, USA
\skiplinehalf
\supit{c} School of Physics \& Astronomy, Cardiff University, 5 The Parade, Cardiff, CF24 3AA, UK
\skiplinehalf
\supit{d} Jet Propulsion Laboratory, 4800 Oak Grove Drive, Pasadena, CA 91109, USA
\skiplinehalf
\supit{e} Optics Department of the Complutense University of Madrid, C.\ Arcos de Jal\'{o}n N\textsuperscript{\underline{o}} 118,\ 28037 Madrid, Spain
\skiplinehalf
\supit{f} Department of Astronomy \& Steward Observatory, 933 North Cherry Avenue, Rm.\ 204, Tucson, AZ 85721, USA
}
}

\authorinfo{Further author information: E-mail: attila@submm.caltech.edu}

\begin{document}

\maketitle

\begin{abstract}
SuperSpec is a pathfinder for future lithographic spectrometer cameras, which promise to energize extra-galactic astrophysics at (sub)millimeter wavelengths: delivering 200--500\,km\,s$^{-1}$ spectral velocity resolution over an octave bandwidth for every pixel in a telescope's field of view. We present circuit simulations that prove the concept, which enables complete millimeter-band spectrometer devices in just a few square-millimeter footprint. We evaluate both single-stage and two-stage channelizing filter designs, which separate channels into an array of broad-band detectors, such as bolometers or kinetic inductance detector (KID) devices. We discuss to what degree losses (by radiation or by absorption in the dielectric) and fabrication tolerances affect the resolution or performance of such devices, and what steps we can take to mitigate the degradation. Such design studies help us formulate critical requirements on the materials and fabrication process, and help understand what practical limits currently exist to the capabilities these devices can deliver today or over the next few years. 
\end{abstract}

\section{INTRODUCTION}

The recent bounty of the {\em Herschel Space Telescope} includes the discovery of around a quarter million submillimeter and infrared bright galaxies. However, while finding submillimeter-selected galaxies (SMGs) is relatively straightforward, studying them has proved extremely challenging, because ({\em a}) they are faint at other wavelengths and ({\em b}) there are often multiple possible optical/NIR counterparts \cite{Pope2006, Younger2009b} within the typical 10''--30'' resolution of 10-m class or smaller telescopes operating in the (sub)millimeter regime.

Redshift determination and cross-identification are currently possible for the brightest (sub)millimeter sources only. Within the (sub)millimeter range, where cross-identification is least problematic, rotational transitions of CO have been used for a few exceptionally bright sources at the IRAM 30-m telescope \cite{Weiss2009a} or the Plateau de Bure Interferometer (PdBI)\cite{Cox2011}, and with the grating spectrometer Z-spec \cite{Z-spec}. Redshifted C$^+$ was detected by ZEUS\cite{ZEUS}. None of these instruments however are suited for the fainter, more typical SMGs. Even if they might allow studying the average properties of the fainter population through statistical means, we know almost nothing about the physics and environment of the individual galaxies that dominate the {\em Herschel} surveys, and account for the bulk of the star-formation activity in the universe. 

What astronomers really need is a powerful multiobject spectrometer in the (sub)millimeter regime observing $\sim$100 telescope beams or more simultaneously, each of them providing $\mathcal{R}$$\ge$300 spectral resolution over an octave bandwidth. Such redshift machines will be used to search for CO or C$^+$ transitions, and could be deployed on cryogenic space telescopes like SPICA\cite{SPICA} or CALISTO/SAFIR, and on large ground based facilities like CCAT\cite{CCAT}. A 100-pixel spectrometer on CCAT, operating in two bands (e.g.\ CCAT bands 1 \& 2), would be around 15 times(!) faster than ALMA in conducting redshift surveys of the sub-millimeter fields.

Imaging spectrometer cameras can also study critical gas coolants in the ISM (e.g.\ CO, C$^+$, Si$^+$), the UV-fields via fine-structure lines (e.g. O$^{++}$, Ne$^{++}$, N$^{++}$, S$^{++}$, and N$^{+}$, Ne$^{+}$), and provide a tomography of reionization, and the star-forming universe using these transitions.


A moderate resolution spectrometer with $\mathcal{R} \sim 600$ is well-matched to the typical $\sim$500\,km\,s$^{-1}$ linewidths of typical active galaxies (at the same time $\mathcal{R} \sim 300$ would be useful, while $\mathcal{R} \sim 1500$ would enable kinematic studies also). For example, a spectrometer covering the 1\,mm atmospheric window (190 -- 320\,GHz) would cover at least at least one CO rotational transition at all redshifts (two or more transitions above z=0.8!), or the bright 158\,$\mu$m C$^+$ hyperfine line at 4.9$<$z$<$8.7. 

SuperSpec\cite{shirokoff,barry} is such a spectrometer, based on ultra-compact thin-film lithography, representing around 100-fold shrinking in all dimensions, and a weight reduction of $\sim$6 orders of magnitude, when compared to current state-of-the-art spectrometers with similar specs. We aim to build and demonstrate a single-pixel pathfinder device for the Caltech Submillimeter Observatory (CSO) within 2--3 years. This work evaluates the feasibility of our novel approach outlined by Kov\'{a}cs and Zmudzinas in December 2011 (Caltech memo), and identifies some of the challenges that we must meet before the technology can supply the large-format (100-pixel to a kilopixel) sub-millimeter spectrometer cameras of the future. 

The structure of this paper is as follows: Section~\ref{sec:superspec} introduces the concept of SuperSpec. Section~\ref{sec:simulations} describes the simulations that provide the analysis of this work. In Section~\ref{sec:discussion}, we validate the concept and discuss some of the limiting aspects to this approach (esp.\ the effect of losses and tolerances), and what steps we can take to mitigate the loss of performance. Finally, we summarize our conclusions in Section~\ref{sec:conclusions}.

\section{SUPERSPEC: A TRANSMISSION-LINE SPECTROMETER}
\label{sec:superspec}


\begin{figure}
  \centering
  \includegraphics[width=0.45\textwidth]{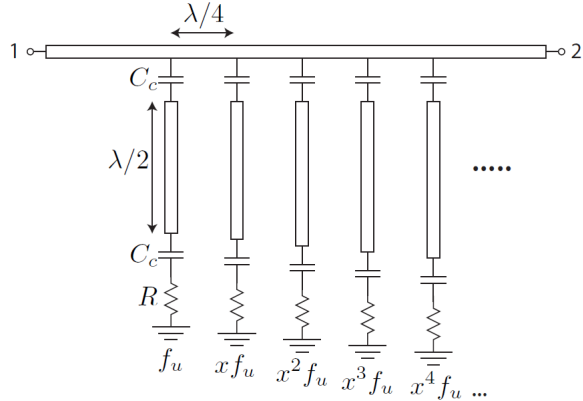}
  \caption{A schematic of SuperSpec. Radiation enters from the left. A sequence of tuned resonators, with logarithmically increasing wavelengths, couple individual narrow sub-bands into power detectors (shown here as resistive terminations), such as bolometers or kinetic inductance detectors (KIDs). The spectrometer channels are spaced along the feedline for optimum efficiency.}
  \label{fig:schematic}
\end{figure}

The concept of SuperSpec constructs a millimeter-wave spectrometer from a collection of narrow-band electronic filters rather than relying on 2D or 3D optical interference (such as grating spectrometers, or Fabri-Perot spectrometers), which inherently limits size to at least $\mathcal{R} \lambda$ along 2 or 3 dimensions. The idea of a spectrometer comprised of channelizing filters is not new and has been used for filterbank backends in radio astronomy in the 1990s\cite{filterbank}. In our design, a spectrometer channel is constructed from a half-wave transmission-line resonator coupled weakly to the input on one end and to a power detector at the other end. The strength of the coupling detemines the quality of the resonator ($Q = f / \Delta f$), and therefore the resolution of the spectral channel ($\mathcal{R} = Q$). 

The novelty of SuperSpec, and of the conceptually similar DESHIMA\cite{akira-jltp, akira-spie}, is in the adaptation to thin-film lithography. At (sub)millimeter wavelengths, transmission lines are implemented either as microstrips or as coplanar waveguides (CPW). Weak reactive coupling is usually by proximity between transmission lines. Last, but not least, the signal can be distributed to all channels along a single input feedline, marking the most important difference compared with traditional filterbanks, where channels are typically connected to a common node. 

When two consecutive channels with overlapping resonances around wavelength $\lambda$ are spaced at intervals of $\lambda/4$ from one another along the feedline, much of the rejected input from given channel is either absorbed by a neighboring channel, or reflected from it and re-arriving in-phase once again. Thus, while an optimally coupled single resonator can absorb at most 50\% of the incident power on resonance from the feed, the $\lambda/4$ spacing of channels, with monotonically decreasing (or increasing) frequencies, can push up the efficiency arbitrarily, even if neighboring channels operate at slightly offset frequencies from one another.

Because of the small size of lithographic transmission lines (microstrip lines can be made reliably as small as 1\,$\mu$m wide with the deep-UV process at JPL, while 'narrow' CPWs can be $\sim$10\,$\mu$m across), it should be possible to produce a spectrometer that is equivalent to Z-Spec\cite{Z-spec} ($\mathcal{R} \sim 250$) or better (e.g.\ covering the octave-wide 1\,mm window with $\mathcal{R} \sim 300$--$1500$), but occupies just a few mm$^2$ of deposition layers on a chip instead of the ca.\ 60\,cm $\times$ 60\,cm metal waveguide enclosure of Z-Spec. 

A lithographic filterbank therefore results in at least a 100-fold reduction in all dimensions, and hence a $\sim$6 orders of magnitude shrinking of the cooled-mass for the entire spectrometer when compared to current state-of-the-art instruments like Z-Spec or ZEUS\cite{ZEUS}. Provided that the required number of sensitive detectors can be packed in a similar volume, this new technology opens the tantalizing possibility of a complete millimeter-wave extra-galactic spectrograph with an octave bandwidth realized on an $F \lambda$ focal-plane pixel, and therefore the advent of fully sampled focal plane spectrometer arrays in the not too distant future. We believe that a 100-pixel to a kilopixel R$\sim$600 array for CCAT is realistic by 2020, promising to deliver a powerful multiobject mapping spectrograph and redshift machine for the (sub)millimeter-bright galaxy (SMG) population. Such devices will be optimal for deployment in space (especially on cryogenic telescopes like SPICA\cite{SPICA} or CALISTO/SAFIR), on balloon missions, and on airborne observatories also, due to their small size/mass, and the correspondingly modest requirement for cooling power.



 


The basic design procedure for a simple resonator is described by Kov\'{a}cs \& Zmuidzinas (2011, Caltech memo). By changing the capacitive couplings at the two ends of the resonator we can dial-in the desired value for the quality $Q_r$ of the resonator. In the zero coupling limit (infinite $Q_r$), corresponding to perfect open ends, the resonator line is exactly half-wave long. As the coupling increases in strength (i.e.\ $Q_r$ decreases) the end reflection shifts slightly in phase by $\delta \Theta$, where $\tan(\delta \Theta) \propto C \omega$, requiring a corresponding correction to the resonator length by $\delta \Theta_r = - \delta \Theta / 2$ at both ends. Thus, optimally-sized resonators are generally shorter than half-wave. One of the aims of this work is to measure the scaling relations for the precise sizing of coupling capacitors and resonator lengths as a function of frequency.




\section{SIMULATIONS}
\label{sec:simulations}

\begin{figure}
  \centering
  \includegraphics[width=0.95\textwidth]{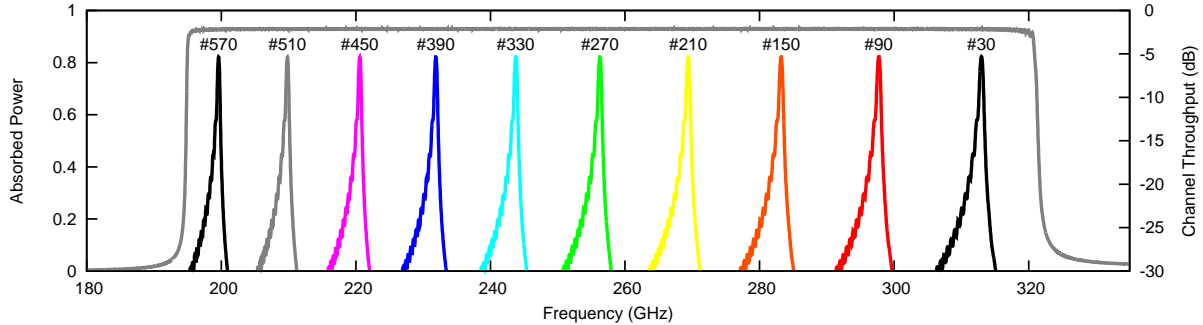}
  \caption{The incident power fraction coupled to spectrometer channels (gray line, left axis) for an $\mathcal{R}=600$ ideal spectrometer with 600 channels, covering the entire 1\,mm atmospheric window. A sampling of 10 individual channel profiles are also shown (right axis).}
  \label{fig:combo-opt}
\end{figure}

We used the SuperMix \cpp\ library \cite{supermix}, developed at Caltech, to simulate an entire spectrometer composed of a large number of channelizing filters, covering an octave-wide band with a spectral resolution $\mathcal{R}$ between 300 and 1500. SuperMix uses a circuit model, with various typical circuit elements connected through nodes, to calculate the full scattering matrix of the device for its open ports. The main advantage of this approach is its speed when compared to finite-element E-M analysis in 2D or 3D\cite{barry}. An entire $\mathcal{R}=600$ spectrometer, comprised of 600 channels, can be evaluated at a thousand different frequencies in about 1 minute on a typical PC, compared with days or even months that would be required for a full-blown E-M simulation for a device of such size and complexity. 

One must note, however, that the circuit model has its inherent limitations. For example, it cannot easily account for proximity coupling distributed over a section of a transmission line (such as we would actually have between lithographic lines). Instead, perfect capacitive couplings between two well-defined points must be assumed. Nor does SuperMix have the ability to estimate radiative losses, or the radiative coupling of the elements (through air, or through a dielectric). In this sense, the SuperMix model is simplistic, without capturing the complexity of the true 3D structure of an actual implementation. Nevertheless, SuperMix does provide a useful approximation of what we may expect, as long as these caveats represent but a small perturbation over the idealized circuit model. 

\begin{figure}
  \centering
  \subfigure[] {
    \includegraphics[width=0.35\textwidth]{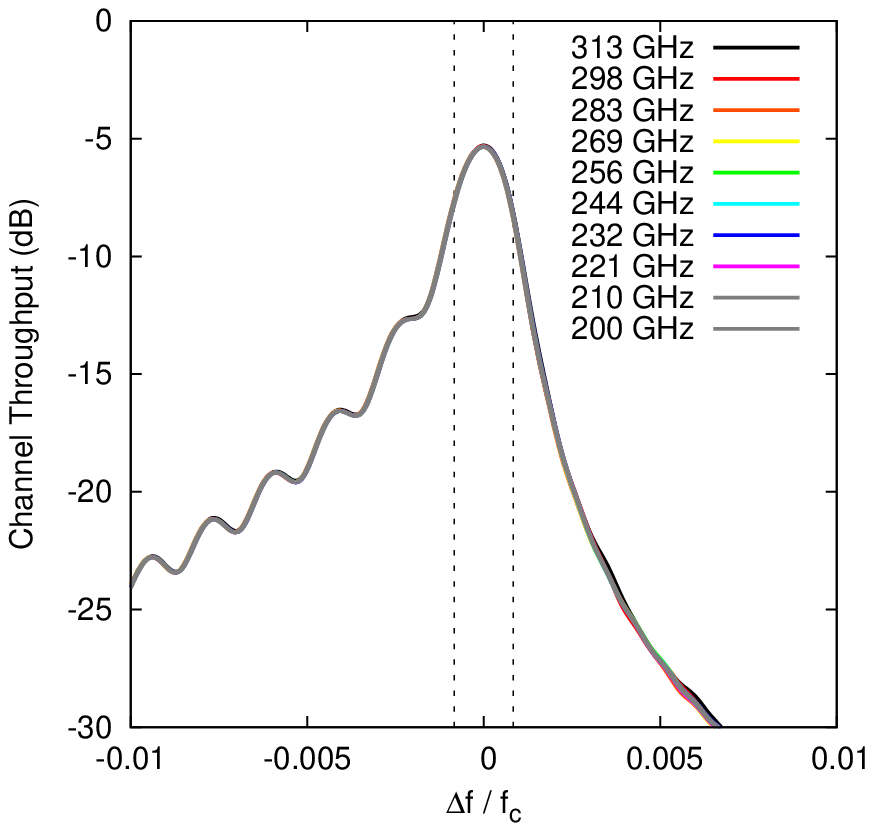} 
    \label{fig:channels-opt}
  }
  \subfigure[] {
    \includegraphics[width=0.35\textwidth]{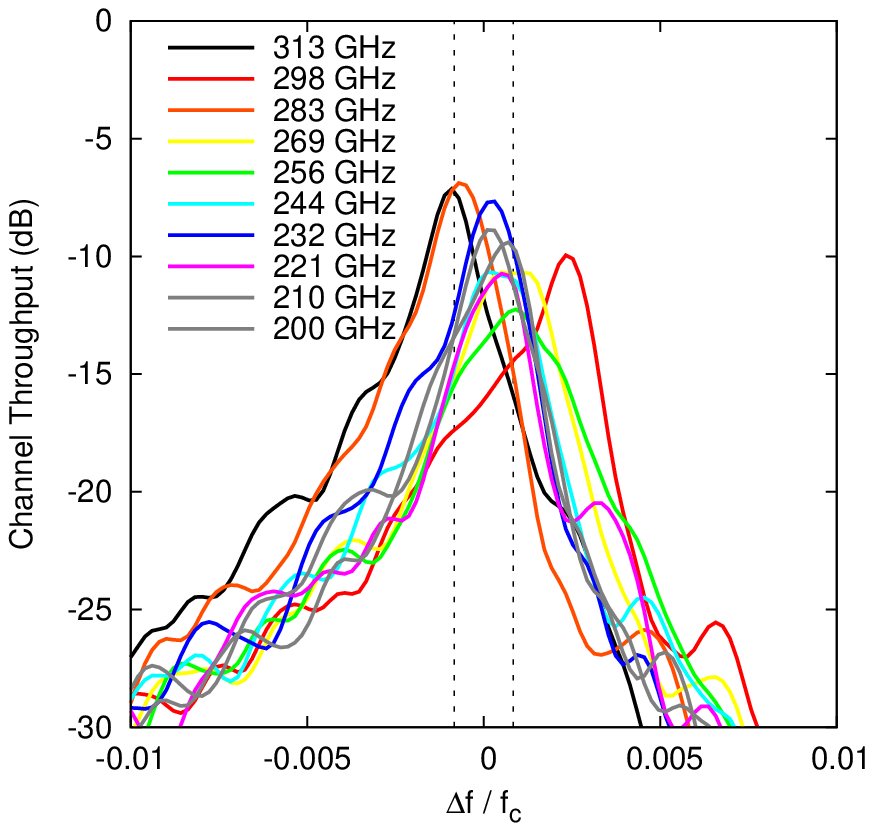} 
    \label{fig:channels-exp}
  }
  \caption{(a) An overlay of the channels shown on Fig.~\ref{fig:combo-opt} exhibit highly uniform spectral profiles across the full band, to the point where they are practically indistinguishable from one-another. The half-power channel bandwidth, corresponding to the spectral resolution $\mathcal{R}$ is show with dotted lines. (b) The same overlay when introducing loss ($\delta \sim 5 \times 10^{-4}$, i.e.\ $Q_{\rm loss} \sim 2000$), a 1\% scatter in the relative resonator couplings channel-to-channel, and a $\sim$0.2\,$\mu$m rms lithographic error between nearby resonators.  }
\end{figure}

Thus, a point-to-point coupling via a perfect capacitor, as modeled by SuperMix, is a generally acceptable model for all reactive coupling (including inductive coupling, provided it is at the proper corresponding location --- e.g.\ for a series capacitance at the end of the resonator with a current node, the corresponding inductive coupling must be at the center of the resonator where there is a voltage node). Transmission-line losses can be included in the SuperMix model if known from elsewhere. Cross-coupling between channels (via air or substrate) can be kept low with careful design (e.g.\ by spacing resonators, using microstrip lines with more constrained fields, or with strategically placed grounds) such that it is not expected to change the spectrometer performance significantly. Last, but not least, all these effects can be studied and characterized by localized small-scale E-M simulations\cite{barry} (of just one or a few filter channels, or parts thereof), and the results can be plugged back into SuperMix for an accurate simulation of the full spectrometer.

In our simulations we assumed a 50\,$\Omega$ feedline (matched to a 50\,$\Omega$ input port, and another 50\,$\Omega$ through port or feedline termination at the other end), 100\,$\Omega$ resonator lines, connecting to 50\,$\Omega$ detectors --- representing an arbitrary choice of parameters. The simulated spectrometers cover nearly an octave bandwidth centered around 250\,GHz (195--321\,GHz), spanning the entire 1\,mm atmospheric window, and matching the coverage of the pursued first-light device for the CSO. Most of the simulations, also assume an spectral sampling density $\Sigma=2$, and a channel count $N$ that equals the targeted resolution ($N = \mathcal{R}$).


\section{DISCUSSION}
\label{sec:discussion}

The simulations clearly show that a spectrometer built according to the principles outlined in Section~\ref{sec:superspec} can couple narrow $\mathcal{R} \sim 300$--$1500$ sub-bands into power-detectors effectively, and with a high-level of uniformity over an octave (or more). Figure~\ref{fig:combo-opt} shows the performance calculated for an ideal device comprised of 600 channels, providing $\mathcal{R}=600$ spectral resolution. As predicted the interference between overlapping neighbors acts only to increase the efficiency of the coupling to $\sim$93\% for a sampling factor $\Sigma=2$. Figure~\ref{fig:channels-opt} testifies to the uniformity of channel responses and $Q$ values across the entire spectrometer band for an ideal system. Fig.~\ref{fig:channels-exp} shows the same for a more realistic simulation that includes losses and random variations (see further details in the sections below). The asymmetry in the profiles arises because the signal at higher frequencies arrives already depleted by the group of nearby channels closer to the input, which couple at these frequencies. (Note, that resonators are arranged from high to low frequencies along the feed.) On the low-frequency side, we can similarly explain the appearance of tiny ripples (around $\pm$1\,dB) with negligible consequence to the performance of the spectrometer as a whole.

\begin{figure}
  \centering
  \subfigure[] {
    \includegraphics[width=0.31\textwidth]{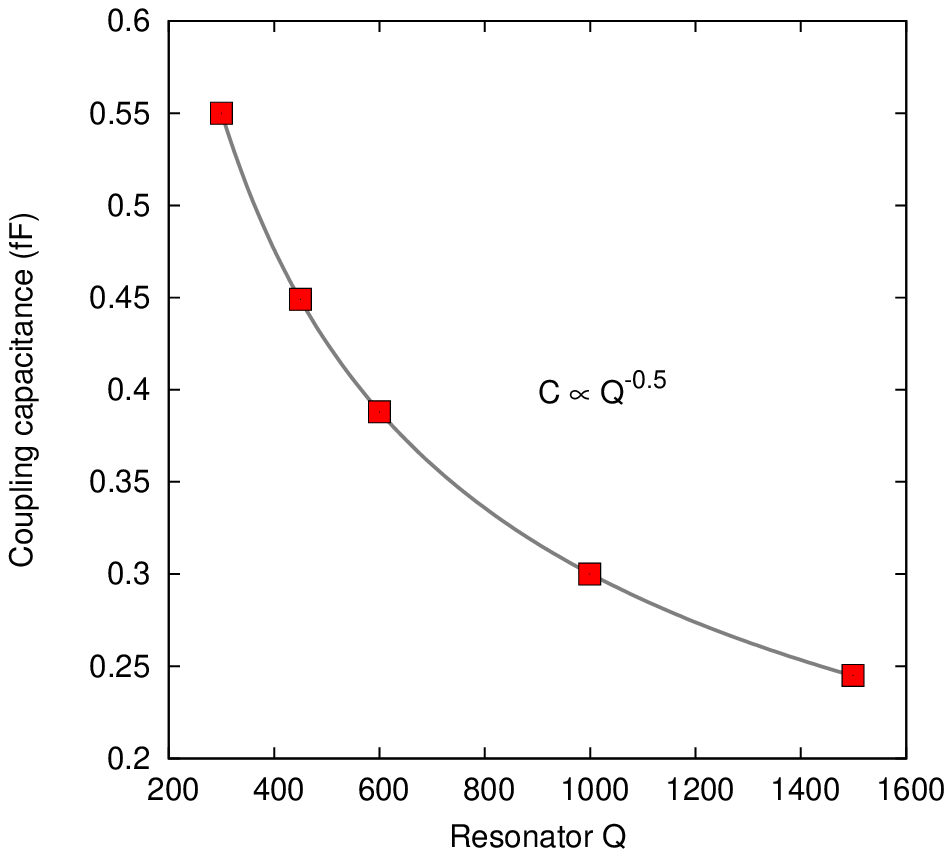}
    \label{fig:Q-C}
  }
  \subfigure[] {
    \includegraphics[width=0.31\textwidth]{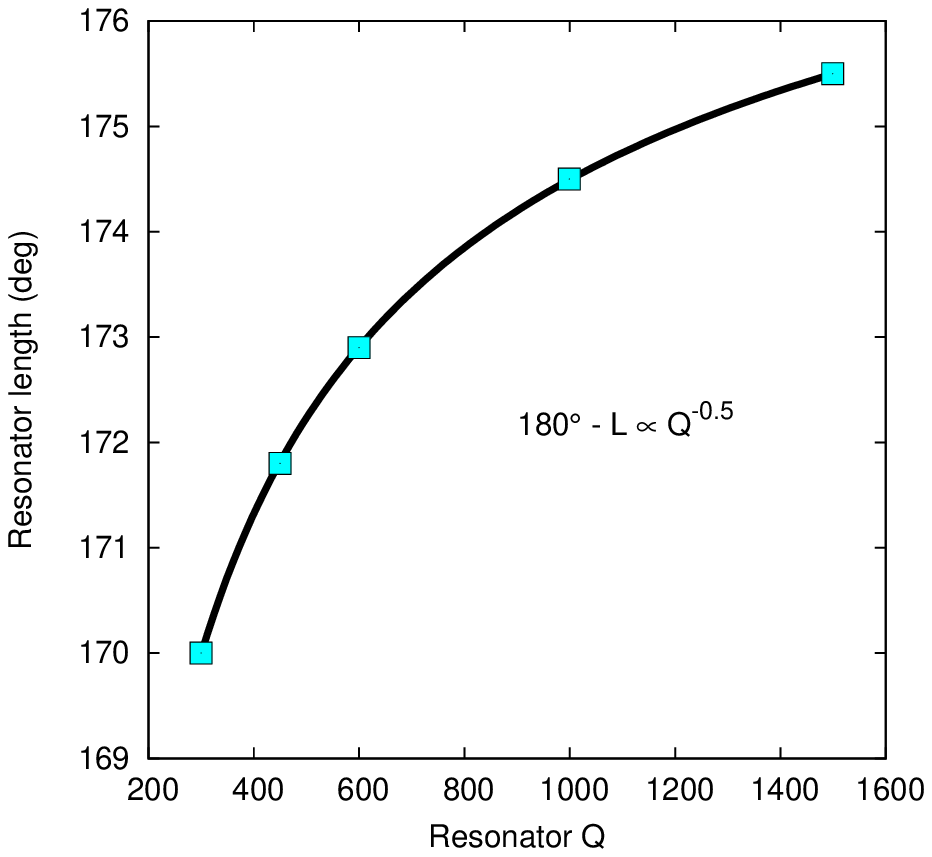}
    \label{fig:Q-l}
  }
  \subfigure[] {
    \includegraphics[width=0.303\textwidth]{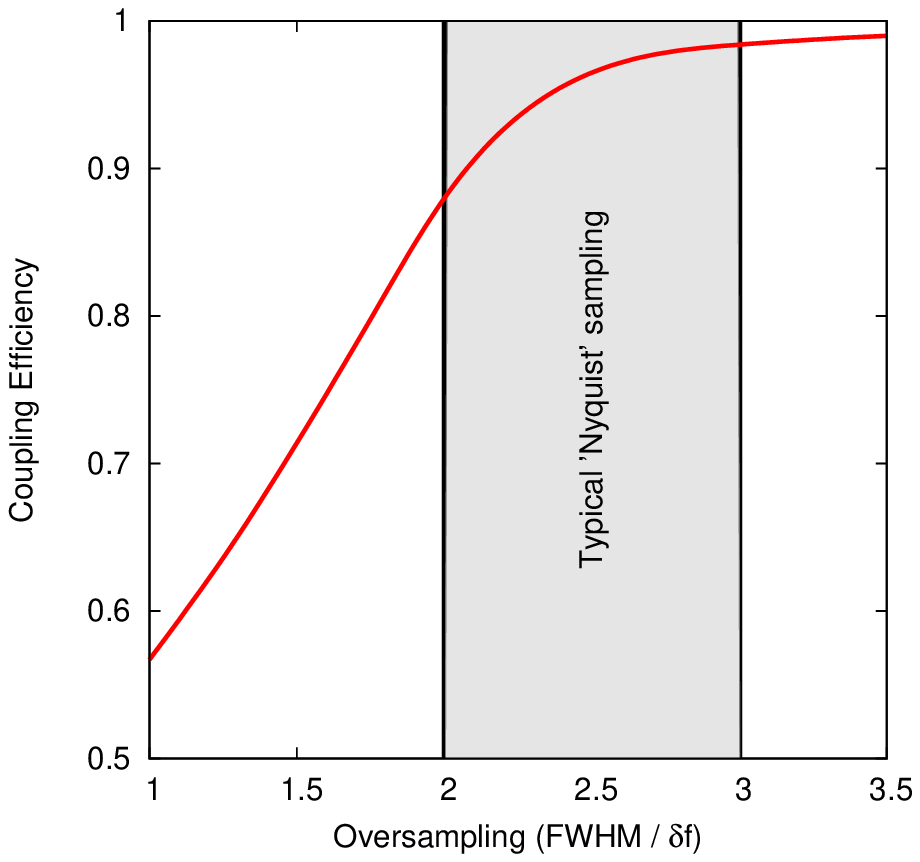} 
    \label{fig:oversampling}
  }
  \caption{(a) The optimal size of capacitive couplings at both ends of the resonator, and (b) the optimal sizing of 'half-wave' resonators to yield the desired $Q_r$ values and center frequencies. Also shown are the best fit models to these, in general agreement with predictions by the Kov\'{a}cs \& Zmuidzinas memo (Caltech, 2011). (c) Shows the power coupling efficiency with spectral sampling density.} 
\end{figure}

Figures~\ref{fig:Q-C} and \ref{fig:Q-l} show the dependence of resonator parameters on $Q_r$ when optimized numerically for the given center frequency. Accordingly both the capacitive couplings and the corrections to the resonator lengths scale with $Q_r^{-1/2}$. This scaling relation has been expected from the theory outlined by Kov\'{a}cs and Zmuidzinas, although the absolute values are somewhat different (within a factor of 2). Therefore, the results gives us confidence that our theory is basically correct, and we will use these data to refine our theoretical understanding in the future. 

We also note, that the simulations verify the need for terminating the excess power at the end of the feedline to avoid harmful reflections. Such reflections can disrupt the carefully constructed interference pattern of the $\lambda/4$-spaced of resonators, and result in non-uniformities at the $\pm$3\,dB level (peak-to-peak) in the coupling efficiency. A practical termination can be into a broad-band power detector (such as a KID or bolometer) providing a useful diagnostic of the out-of-band incident power on the chip.

\subsection{Spectral sampling}

We define the spectral sampling density $\Sigma$, as the ratio of channel bandwidth to the frequency spacing of channels. Figure~\ref{fig:oversampling} shows that the coupling efficiency quickly increases from the 50\% maximum value for an isolated resonator, to 90\% and above as $\Sigma$ is increased to around 2 or above, provided the resonators are also spaced at $\lambda/4$ intervals along the feed. Beyond the efficiency of coupling, information theory also commends $\Sigma$ values around 2 or higher. The Nyquist sampling theorem states that information contained in a signal that is perfectly filtered above a cutoff frequency $f_c$ can be fully reconstructed from a set of discrete samples provided the samples are collected at a frequency strictly higher than $2 f_c$. In reality there never is a perfect low-pass filter, or one with a well-defined frequency cutoff. Instead filters roll off smoothly (some perhaps faster than others) at the higher frequencies, blurring the exact interpretation of the Nyquist condition. 

In case of a spectrometer the sampling domain is in frequency, where the information is restricted to a resolution $\mathcal{R}$ with the tapered profile of our resonators, the Fourier pair of which will roll-off slowly also. Choosing a sampling condition of $\Sigma=2$ comprises an approximate Nyquist criterion, where some large fraction (typically $\sim$90\%, depending on the actual channel profile) of the information about the underlying spectrum is preserved, but the remaining part ($\sim$10\%) is aliased, corrupting the information content. One can preserve the underlying signal, of inherent spectral resolution $\mathcal{R}$, more completely and with higher fidelity (less corruption) by increasing $\Sigma$ beyond 2. The trade-off is between more perfect information content (i.e.\ delivering higher-quality spectra) with better coupling vs.\ a larger number of detectors required for the same total bandwidth.

The combination of practical considerations, for device coupling efficiency and for detector counts, and theoretical considerations for spectral fidelity, all converge towards useful spectral sampling densities around $\Sigma \sim 2$--3. Our relatively strong preference for keeping detector counts low explains our choice of $\Sigma=2$ for our initial designs for a first-light SuperSpec implementation for the CSO.

\subsection{Transmission-line losses}
\begin{figure}
  \centering
  \subfigure[] {
    \includegraphics[width=0.37\textwidth]{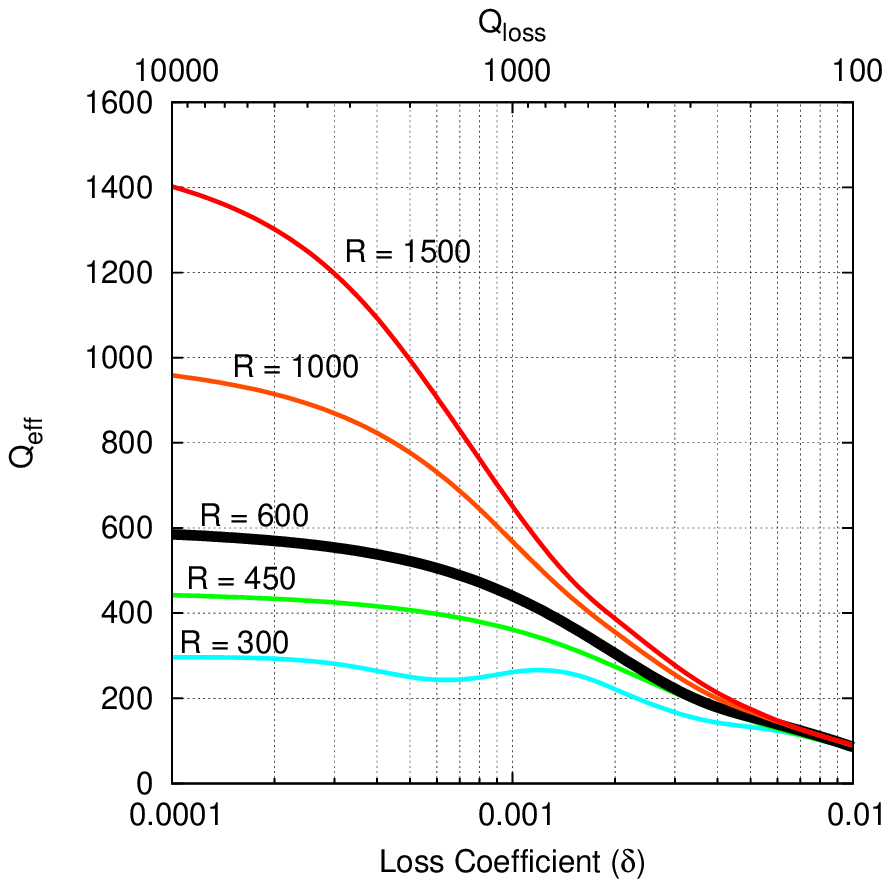}
    \label{fig:Qloss}
  }
  \subfigure[] {
    \includegraphics[width=0.35\textwidth]{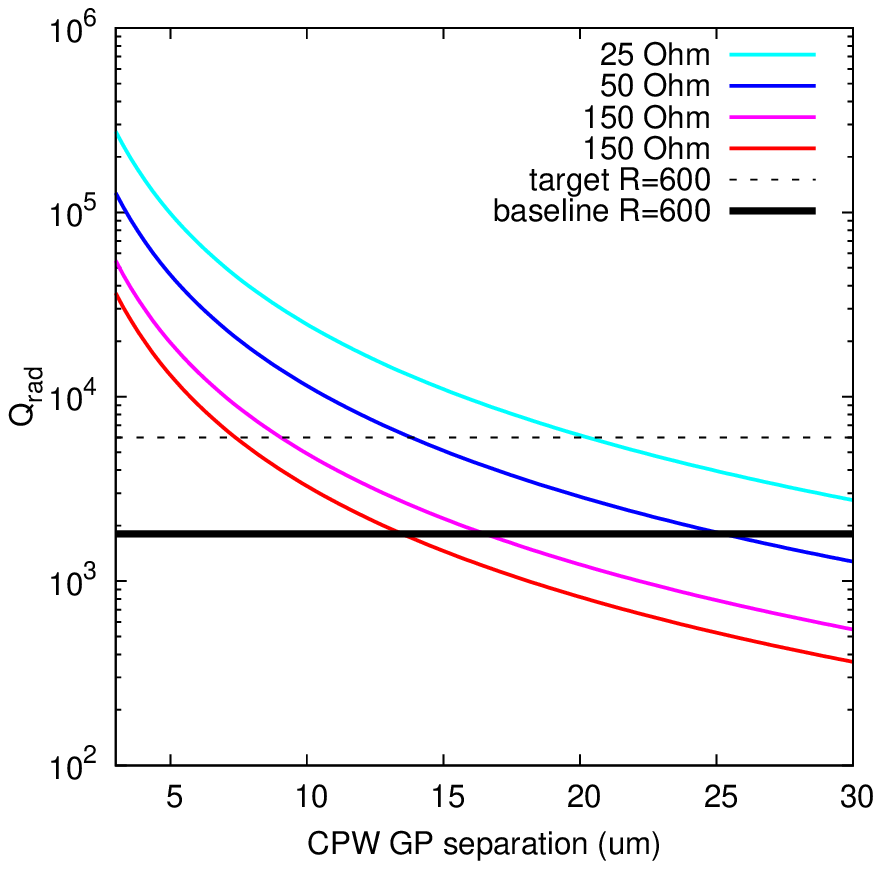} 
    \label{fig:cpwloss}
  }
  \caption{(a) The dependence of the effective resolution of the spectrometer vs.\ the loss coefficient $\delta$ (or equivalently $Q_{\rm loss}$) from the numerical simulations for a number of different target resolutions (here $\mathcal{R} = Q_r$). We identify the asymptotic behavior of these curves for both the low-loss and the loss-dominated regimes. (The slight wiggle in the $\mathcal{R}=300$ case is probably a numerical artifact.) (b) The radiation loss ($Q_{\rm loss} = Q_{\rm rad}$) vs.\ CPW ground-plane separation for different CPW impedances, as predicted by the Vayonakis \& Zmuidzinas memo (Caltech, 2011).}
\end{figure}

One should expect that losses in the transmission lines degrade the effective quality of the resonators, since the some of the signal is dissipated before it travels the typical $\mathcal{R} \lambda$ distance in the resonator. The loss in a transmission-line can be characterized either as a loss tangent $\tan \delta$ (the ratio of the impedance in the reactive direction to the impedance in the dissipative direction) or as $Q_{\rm loss} \equiv 1/\delta$. 

In this work we do not offer an expression for the effective quality $Q_{\rm eff}$ of a resonator in terms of its intrinsic $Q_r$ and the loss. Instead, we relied on the simulations to characterize this effect empirically. We find that the $Q_{\rm eff}$ of a lossy resonator varies as 

\begin{equation}
Q^{-1}_{\rm eff} \approx Q_r^{-1} + Q_{\rm loss}^{-1}
\label{eq:Qloss}
\end{equation} 

when losses are low ($Q_{\rm loss} \gg Q_{r}$) and as $Q^{-1}_{\rm eff} \approx Q_r^{-1} + 1/2~Q_{\rm loss}^{-1}$ when losses become dominant ($Q_{\rm loss} < Q_{r}$). 

To control the effective resonator $Q$s (and with it the resolution of the spectrometer) we want to be in the regime dominated by the intrinsic quality of the resonator. Accordingly, $Q_{\rm loss}$ ought to exceed $Q_r$ by a factor of $\sim$3 the least, and ideally by an order-of-magnitude or more. Then, the effective quality is adequately described by the asymptotic form of Eq.~\ref{eq:Qloss}, which can be used for estimating the intrinsic resonator quality $Q_r$ necessary for maintaining a spectral resolution $\mathcal{R} = Q_{\rm eff}$ with lossy transmission lines. 

In microstrip lines, the loss mechanism is predominantly dissipation in the dielectric layer, which depends on the material properties. A 250\,GHz loss tangent of around 5$\times$10$^{-4}$ or lower, corresponding to $Q_{\rm loss} \sim 2000$ or higher, should be possible with SiN or amorphous Si substrates\cite{Martinis2005}.

The loss in CPW lines is due to radiation in the slotline mode. Vayonakis \& Zmuidzinas (2012, Caltech memo) approximate the radiation loss in a half-wave CPW resonator in terms of its geometry. In Figure~\ref{fig:cpwloss} we show their estimates for a number of CPW impedances and ground-plane separations at 250\,GHz. Accordingly, the radiation loss in CPW resonators should enable $\mathcal{R} \sim 600$ spectroscopy for the millimeter band as long as ground-plane separations remain below $\sim$10\,$\mu$m. Ideally, we would like to validate the expressions by Vayonakis \& Zmuidzinas, e.g.\ by comparison to numerical measures of radiation loss in 3D E-M simulations\cite{barry}, or via direct measurement.


\subsection{Tolerance analysis}

\begin{figure}
  \centering
  \subfigure[] {
    \includegraphics[width=0.3\textwidth]{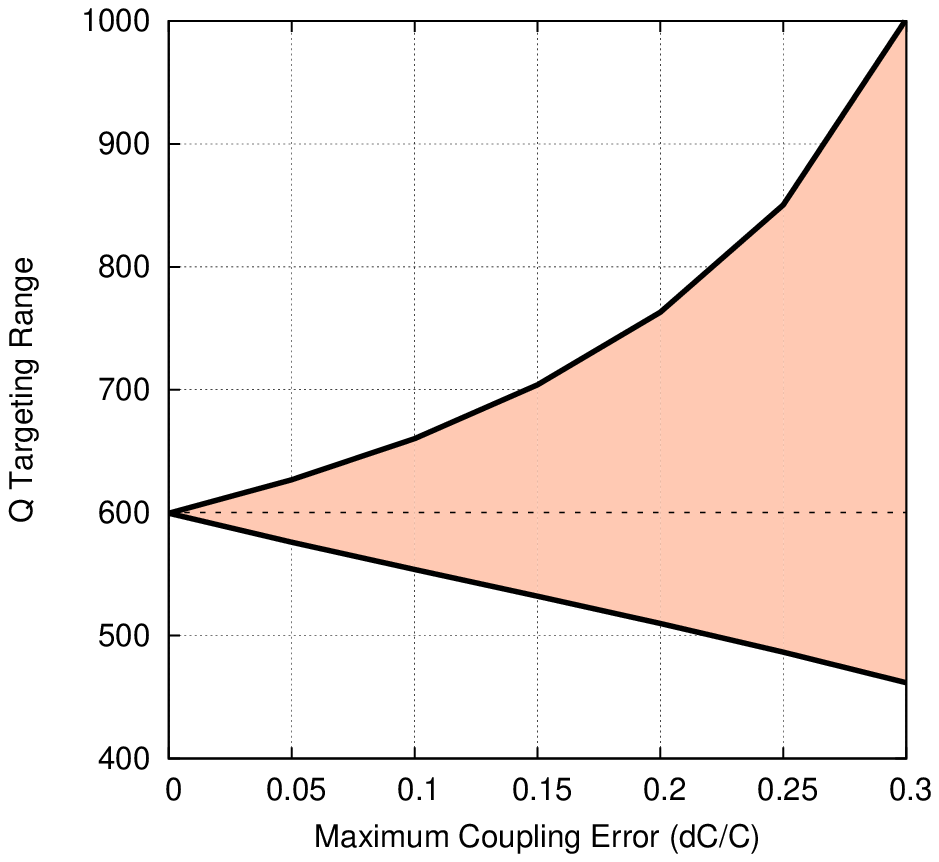}
    \label{fig:Cerr-dQ}
  }
  \subfigure[] {
    \includegraphics[width=0.3\textwidth]{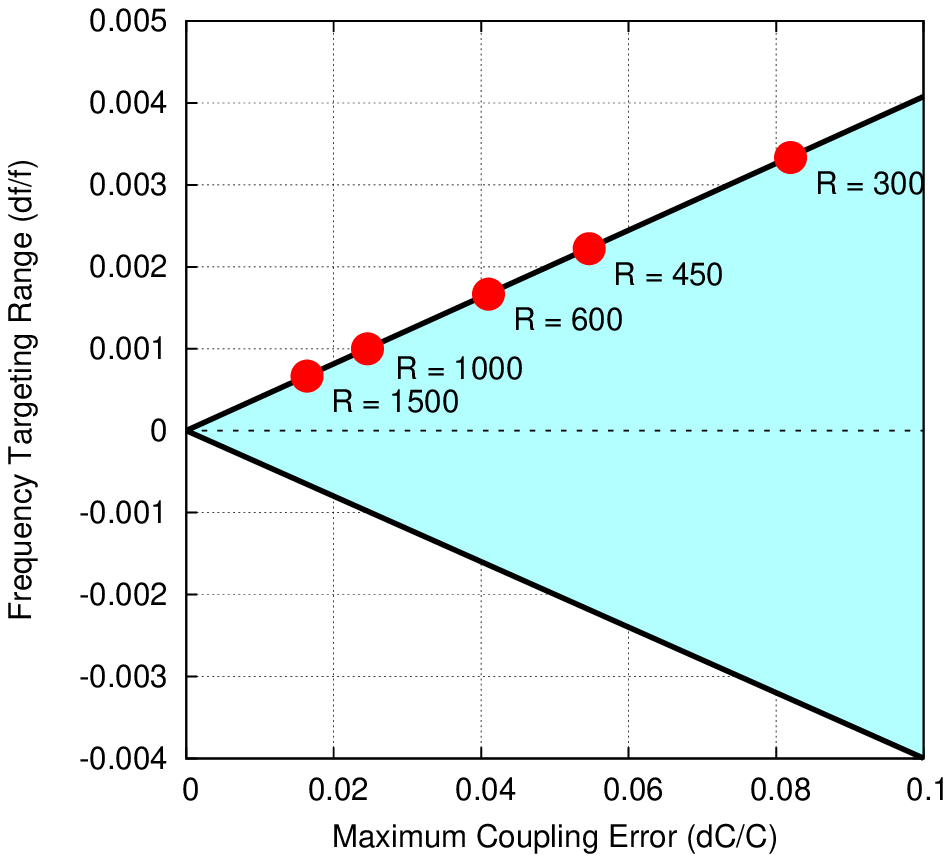}
    \label{fig:Cerr-df}
  }
  \subfigure[] {
    \includegraphics[width=0.3\textwidth]{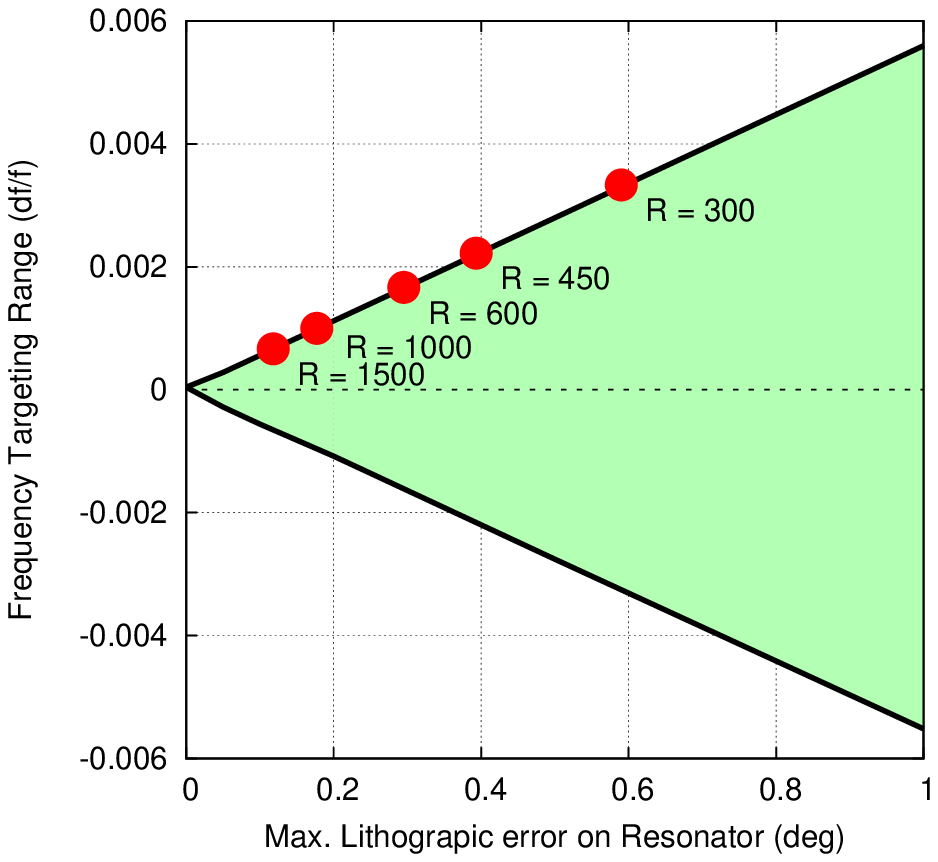} 
    \label{fig:lerr}
  }
  \caption{(a) The range of variation in the spectral resolution of channels ($\mathcal{R} = Q$) vs the maximum fractional error in the coupling strength. Panels (b) and (c) show the range of frequency shift among channels vs.\ the fractional error in the coupling strength and resonator phase length, respectively. The red dots indicate the maximum channel-to-channel error we can tolerate before channels are displaced by more than their widths, relative to one another. Larger errors will result in an unpredictable resonator order.} 
\end{figure}

Fabrication errors further limit the performance of actual lithographic devices. All parameters of the lithography have associated variations: line widths, lengths and separations; layer thicknesses; dielectric properties; and conductor/absorber material properties. In terms of the resonators these translate to variations in resonator lengths, coupling strengths, and line impedances. Here we consider mainly the effect of variations in the coupling strength and in the resonator lengths.

Figures~\ref{fig:Cerr-dQ}, \ref{fig:Cerr-df} and \ref{fig:lerr} show that a targeting error in the coupling affects both $Q$ and the frequency of the resonance, while a phase-length error results primarily in just a frequency shift. Red dots indicate the maximal errors we can tolerate for various $\mathcal{R}$ values before channels shift by more than their widths. Note, that the spacing of channels is smaller than their width by a factor of $\Sigma$, such that some neighboring pairs of channels may switch order even sooner. Nevertheless, to keep channel order predictable, at least to the precision of the resolution of the spectrometer itself, the maximum error in the coupling strength we can tolerate is of order $25/\mathcal{R}$, or around 4\% for $\mathcal{R}=600$. At this level, the expected fractional variation of $Q$ more or less mimics the fractional variation in the coupling. Similarly, the maximum tolerance to resonator lengths is about $180^\circ / \mathcal{R}$ or about 0.30$^\circ$ at $\mathcal{R}=600$. 

For a spectrometer with $\sim$1000 channels, excursions will occur up to around 3$\sigma$ level. Thus, we should aim to keep the fractional variations in the couplings below $\sim$$8/\mathcal{R}$ rms, and the variation of lengths below $\sim$60$^\circ / \mathcal{R}$ rms. When variations exceed these limits, the channel frequencies and responses must be measured and calibrated individually.  
Clearly, these strict limits do not apply to the absolute targeting errors (which will shift all channels by the same amount), nor do they refer to variations on a global scale across the wafer. Rather, the limits apply to relative variations between neighboring channels only. 

Let us translate these numbers into practical terms of a typical lithography: for proximity coupling between two transmission lines coupled at a distance of 5\,$\mu$m, the difference between the line gaps at two neighboring channels need to be maintained to a relative accuracy of $\sim$70\,nm rms or below. Similarly, using free-space propagation as a guide, a spectrometer operating at $\sim$1\,mm wavelength must have resonators whose lengths increase to about 0.3\,$\mu$m rms accuracy from one channel to the next to make an $\mathcal{R}=600$ spectrometer work well. Although challenging, these relative accuracies should be possible to achieve with the deep-UV lithography at JPL, between locations separated by no more than 1\,mm on the wafer.

\subsection{Robust design strategies}

\begin{figure}
  \centering
  \includegraphics[width=0.95\textwidth]{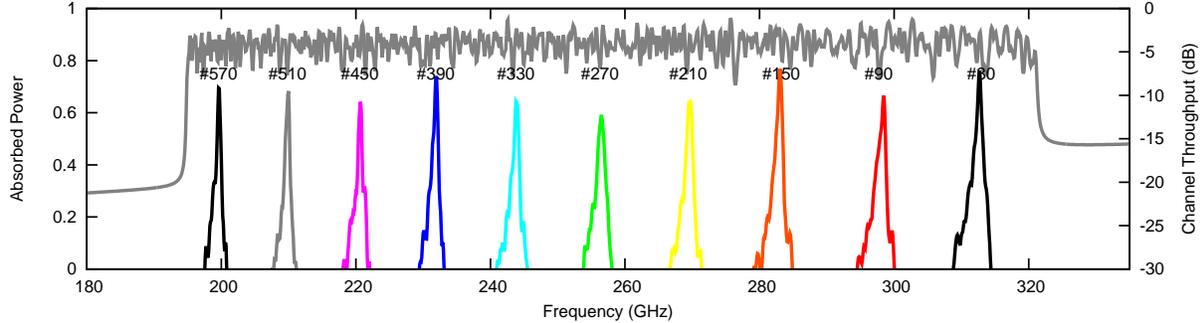}
  \caption{Same as Figs.~\ref{fig:combo-opt}, but including loss ($\delta \sim 5 \times 10^{-4}$ or equivalently $Q_{\rm loss} \sim 2000$), a 1\% scatter in the relative resonator couplings channel-to-channel, and a $\sim$0.2\,$\mu$m rms lithographic error between nearby resonators. These are typical values for what we can expect from the JPL deep-UV process on a SiN substrate. See also Fig.~\ref{fig:channels-exp}.}
 \label{fig:combo-exp}
\end{figure}

As discussed above, both transmission-line losses (radiative or in the dielectric) and fabrication tolerances will degrade performance (throughput, resolution and uniformity) to a certain degree, when compared to the ideal case presented on Figs.~\ref{fig:combo-opt} and \ref{fig:channels-opt}. A typical device may perform as shown on Figs.~\ref{fig:channels-exp} and \ref{fig:combo-exp} instead. However, some of the disruptions can be countered in the design.

Transmission line losses, if characterized, can be offset to a small degree by increasing the internal $Q_r$ of the resonators according to Eq.~\ref{eq:Qloss}. If the channel-to-channel errors in the coupling result in excessive scattering of channel frequencies, such that the frequency ordering of channels is no longer monotonic or predictable, one may increase the gap between coupled lines and couple over a longer section instead to maintain coupling strength overall. A higher sampling density $\Sigma$ increases redundancy among channels, which helps maintain uniform high efficiency absorption of the input even when channels shift around wildly. As a result, a device may be useful even with an unpredictable channel order, if channel frequencies can be calibrated, e.g.\ via an LO sweep.

Last, but not least, targeting a lower spectral resolution $\mathcal{R}$, or lowering the operating frequencies of the spectrometer will help combat losses and tolerance issues alike. Although we hope to build spectrometers with $\mathcal{R} \sim 600$ or higher for the 1\,mm band, a spectral resolving power of $\mathcal{R} \sim 300$ operating in the 2\,mm atmospheric window would already be useful for conducting sensitive redshift searches today. Over time, new research into dielectrics and higher-precision lithography will enable spectrometers with higher resolutions, and/or operating at shorter wavelengths. 

\subsection{Muliple resonator stages}

\begin{figure}
  \centering
  \subfigure[] {
    \includegraphics[height=0.32\textwidth]{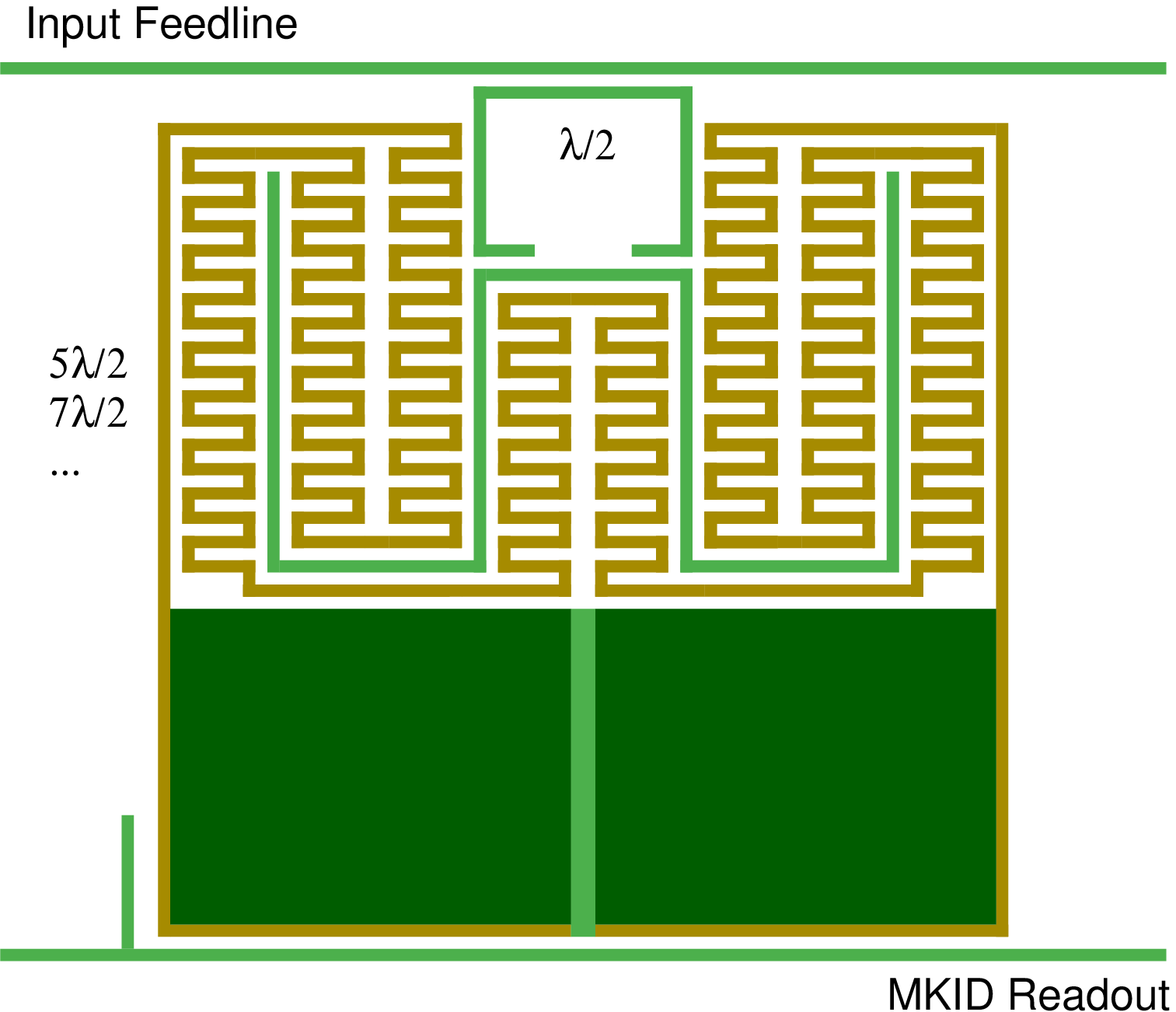}
    \label{fig:twopole}
  }
  \subfigure[] {
    \includegraphics[height=0.32\textwidth]{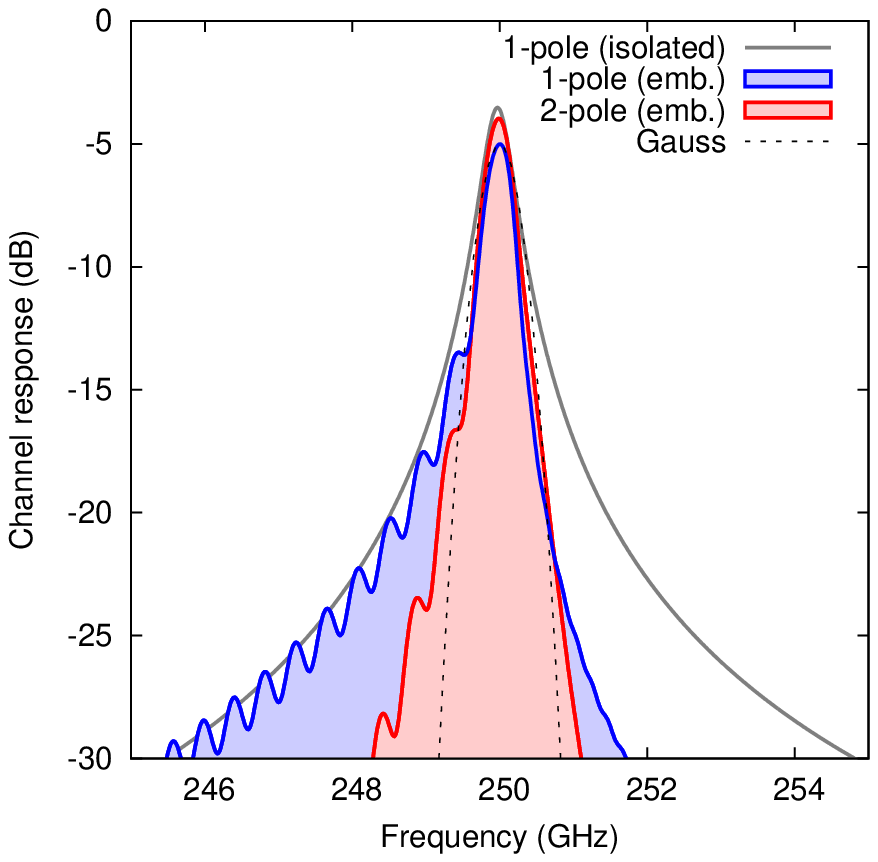} 
    \label{fig:profile}
  }
  \caption{(a) A schematic layout of a 2-stage resonator device coupled to a meandered KID device through a longer second-stage resonator. (b) An overlay of the relative channel profiles of a single half-wave resonator, and a 2-pole resonator, with a 5$\lambda / 2$ second stage. For comparison, we show the response of an isolated half-wave resonator that is not embedded in a spectrometer, and a Gaussian profile also.}
\end{figure}

Thus far, we have considered only a single half-wave resonator for our spectrometer channels, which has the response of a 1-pole bandpass filter. We may, however, combine multiple resonators in series to improve the response of channels. A schematic of a 2-stage resonator, with a second-order response, is shown on Fig.~\ref{fig:twopole}. A similar 2-stage design was recently presented by Endo et al.\cite{akira-jltp}. We may use the higher-order responses either for making the bandpass of the channels flatter, or for providing more isolation between channels (steeper wings) for example. 

Figure~\ref{fig:profile} compares the responses expected from 1-stage and 2-stage resonators. Here, the 2$^{\rm nd}$ stage was tuned for a steeper out-of-band response, thus minimizing the cross-talk between channels. This 2-pole response is almost Gaussian in shape, with properties that are analogous to the Gaussian telescope beams on sky, but in frequency space instead of a spacial domain. Also, the second-stage can be longer, around any odd-half wavelength, $l \sim \ (2 n + 1) \lambda/2$, since the first half-wave resonator will uniquely select the mode from the octave (or more) of available bandwidth. This is especially useful if the power from the resonators needs to be absorbed over a longer distance into the detector (e.g.\ for a KID device, both to create a larger active region, and to decrease the density of quasi-particles). Thus, in our simulations we considered a second resonator, whose length was around $\sim$$5 \lambda / 2$.

One should note, however, that multiple resonator stages have higher complexity and thus place more stringent requirements on the fabrication tolerances, since the improvements mainly rely on the precision tuning between stages, even more than between neighbouring channels since channels profiles are expected to be sensitive to changes at the frational level of a channel width.

\section{CONCLUSIONS}
\label{sec:conclusions}

We demonstrated the feasibility, via circuit simulations, of an ultra-compact on-chip spectrometer using lithographic transmission lines, in the (sub)millimeter bands. We verified that the design recipe put forward by Kov\'{a}cs and Zmudzinas is approximately correct, and produces the expected scaling relations. Our other conclusions are:

\begin{itemize}
\item{It is important to terminate the input feed, e.g.\ by a power detector that can also be used as a diagnostic device.}
\item{A spectral sampling density $\Sigma \sim 2$--3 is both practical and desirable. It helps improve the power coupling efficiency of the spectrometer to $\sim$100\%, and it is needed to preserve spectral information more completely at the intrinsic resolution of the device.}
\item{Both microstrip (e.g.\ on SiN or amorphous Si dielectrics) and CPW resonator implementations should provide $\mathcal{R} \sim 600$ spectral resolution in the 1\,mm window (190--320\,GHz) reliably.}
\item{Although transmission line losses will limit the resolutions we may reach, we can partially compensate by increasing the intrinsic quality $Q_r$ of the resonators.}
\item{We find that the implementation puts stringent requirements on fabrication tolerances: an $\mathcal{R} \sim 600$ spectrometer requires gaps between coupled lines to be controlled to $\sim$70\,nm rms channel-to-channel, and relative lengths of nearby resonators to controlled to below $\sim$0.3\,$\mu$m rms, for keeping the frequency ordering largely undisturbed.}
\item{If channel ordering is not a concern (e.g.\ because it can be measured by an LO sweep), increasing the sampling density $\Sigma$ can ensure a more uniform input power coupling across the band even when channels shift around wildly.}
\item{If tolerances become too challenging to meet, the requirements on these can be relaxed by decreasing the resolving power $\mathcal{R}$, and/or by operating at a longer wavelength. An $\mathcal{R} \sim 300$ spectrometer operated in the 2\,mm atmospheric window would already be useful for conducting limited redshift-searches.}

\end{itemize}

Because of its ultra-compact format, occupying just a few mm$^2$ of a chip, the SuperSpec concept is ideally suited for large-format spectrometer cameras, with 100--1000 focal plane pixels simultaneously measuring galaxy spectra over an octave band in the (sub)millimeter range. It is also ideal for deployment in space, balloon mission, or in airborne observatories, where weight and cooling power requirements are critical considerations. The technology of SuperSpec is thus likely to power many of the great (sub)millimeter surveys of the future.

\subsection*{Acknowledgements}

M.~Hollister and L.~Swenson acknowledge funding from the NASA Postdoctoral Program. C.~McKenney, E.~Shirokoff, and L.~Swenson acknowledge support by the Keck Institute for Space Studies. We also acknowledge funding for the development of SuperSpec by the NASA Astrophysics Research and Analysis grant \# 399131.02.06.03.43. P.\ S.\ Barry acknowledges the continuing support from the Science and Technology Facilities Council Ph.D studentship programme and grant programmes ST/G002711/1 and ST/J001449/1.

\bibliographystyle{spiebib}   





\end{document}